\documentclass[aps,prb,onecolumn,groupedaddress]{revtex4}

\usepackage{graphicx}

\newcommand{\beq}{\begin{equation}}
\newcommand{\eeq}{\end{equation}}
\newcommand{\bdis}{\begin{displaymath}}
\newcommand{\edis}{\end{displaymath}}
\newcommand{\bea}{\begin{eqnarray}}
\newcommand{\eea}{\end{eqnarray}}
\newcommand{\barr}{\begin{array}}
\newcommand{\earr}{\end{array}}
\newcommand{\bfig}{\begin{figure}[!]}
\newcommand{\efig}{\end{figure}}

\begin{document}

\title{Comment on "Compositional and Microchemical Evidence of Piezonuclear Fission Reactions in Rock Specimens Subjected to Compression Tests" [Strain 47 (Suppl. 2), 282 (2011)]}

\author{G. Amato, G. Bertotti, O. Bottauscio, G. Crotti, F. Fiorillo, G. Mana, M. L. Rastello, P. Tavella, and F. Vinai}

\address{INRIM - Istituto Nazionale di Ricerca Metrologica, Strada delle Cacce 91, 10135 Torino, Italy}

\date{Received \today}

\begin{abstract}
it is shown that the chemical composition data published by Carpinteri et al. in the article "Compositional and Microchemical Evidence of Piezonuclear Fission Reactions in Rock Specimens Subjected to Compression Tests" [Strain 47 (Suppl. 2), 282 (2011)] cannot be the result of independent measurements as claimed by the authors. Therefore, no conclusion can be drawn from them about compositional modifications induced in the stone by hypothetical piezonuclear reactions taking place during catastrophic failure of the material at fracture.
\end{abstract}

\maketitle

\section{Introduction}

In the article by Carpinteri et al. \cite{Carpinteri2011}, energy-dispersive X-ray spectroscopy (EDS) was used to investigate the compositional properties of samples of Luserna stone. These samples were  taken from specimens previously subjected to fracture tests to reveal the occurrence of hypothetical piezonuclear reactions \cite{Carpinteri2009,Carpinteri2009a}. The goal was to give EDS evidence that a detectable fraction of Fe nuclei had been transformed into lighter nuclei (e.g., Al) by piezonuclear reactions taking place during the catastrophic failure of the material at fracture.

The EDS compositional analysis was carried out for selected crystalline phases in the Luserna stone, and results pertaining to the same crystalline phase before and after fracture were compared. Two crystalline phases, phengite and biotite, were considered, because of their high Fe content. The present comment concerns the phengite data only. Nothing is said about the biotite data.

In the case of the phengite phase, EDS spectra were collected at 30 sites on polished thin sections of the uncracked external surface, and at 30 sites on small portions of the fracture surface. Semi-quantitative standardless analysis was performed on these 60 spectra, fixing the stoichiometry of the oxides. The results of this analysis  are listed in Table I for the external surface  and in Table II for the fracture surface. These tables are taken from the paper by Carpinteri et al. \cite{Carpinteri2011} with no modification.

Carpinteri et al. \cite{Carpinteri2011} claimed that the data of Tables I and II give clear evidence of reduced Fe content at fracture surfaces, with simultaneous increase of lighter elements (Mg, Al, Si).

In this comment, it is shown that the compositional data listed in Tables I and II cannot be the result of independent measurements. Therefore, no conclusion can be drawn from them about compositional modifications induced in the stone by hypothetical piezonuclear reactions taking place at fracture.

\section{Discussion of compositional data}
A preliminary examination of the data listed in Tables I and II led to an overall impression of some regularity in the data sets associated with different sites. Therefore, it was decided to go through a more careful and systematic examination of the individual data sets. It was found that, despite the fact that the 60 EDS spectra acquisitions must be independent by definition, the 60 resulting compositional data sets exhibit some sort of correlation. This correlation is revealed when the compositional data (it is sufficient to consider the oxide data) are reordered in the way shown in Tables III, IV, and V. According to these tables, discussed in detail below, 28 of the 30 external surface data sets (left side of the tables) as well as 19 of the 30 fracture surface data sets (right side of the tables) display some kind of correlation. Correlated data are those appearing in a single table row (table rows run from A to M across the three tables). Table VI lists the remaining 2 external surface and 11 fracture surface data sets for which no evident correlation with other data sets could be identified.

It is worth stressing that the considerations and conclusions that follow are based on the mere revisitation of the data of Tables I and II, with no elaboration, filtering or manipulation whatsoever. To avoid from the beginning any risk of erroneous replacement or manipulation of data, Tables III, IV, V, and VI were prepared by old-style (that is, with scissors and glue) cutting and pasting of the original data.

To clarify the nature of the correlation present in the compositional data, one can consider for example row A in Table III. Four compositional data sets appear in this row, measured on external surfaces at sites 1, 11, 21, and 28. One can see that the content of Si, Mg, and K oxides coincides for all data sets down to the second decimal figure. The relative precision of semiquantitative standardless EDS analysis is of the order of some percent for homogeneous test specimens with carefully prepared polished surface \cite{Standard}. Data generated on casual surfaces, such as fracture surfaces, are of significantly lower precision with unpredictable variation \cite{Standard}. On the other hand, a simple analysis of the dispersion of the compositional data listed in Tables I and II shows that their standard deviation is of the order of 1 wt $\%$. Therefore, there is no reason why the above mentioned figures for the Si, Mg, and K oxide content should be exactly coincident. Even if one accepts that for some reason the digits expressing the content of Si, Mg, and K oxides may all coincide except for the last decimal figure (by itself a rare event), the conditional probability that under this assumption also the rightmost digit will coincide for each of the three oxides in all 4 data sets is of the order of  $10^{-9}$, if one makes the assumption that the rightmost digit is a random variable with uniform distribution. This probability is so small that one must conclude that the data listed in row A of Table III are not independent. This state of affairs applies for most of the external surface data sets (see Tables III and IV, left side) and for about half of the fracture surface data sets (see Tables IV and V, right side).

An additional outcome of this analysis is the cross-correlation between external surface data and fracture surface data. For example, one can consider the data sets in row E of Table IV, three of which are from external surfaces and one from fracture surfaces.  Apart from the Ti oxide, the content of each of the other oxides has identical decimal part in all 4 data sets. As previously discussed, one must conclude that these data also are not independent. In addition, the content of Al oxide in the fracture surface data set is larger than that in the external surface data sets by exactly 4 units, the content of Fe oxide is smaller by exactly 5 units, and the content of K oxide is larger by exactly 1 unit. Similar considerations apply, with slight modifications, to all the data listed in Table IV.

\section{Conclusions}
The conclusion of the analysis presented in the previous section is that the chemical composition data published by Carpinteri et al. \cite{Carpinteri2011} (Tables I and II) cannot be the result of independent measurements. Consequently, no conclusion can be drawn about compositional modifications induced by hypothetical piezonuclear reactions taking place during catastrophic failure of the material at fracture.

%: Table I
\begin{table}
\vspace{-3.0cm}
\includegraphics[width=19cm]{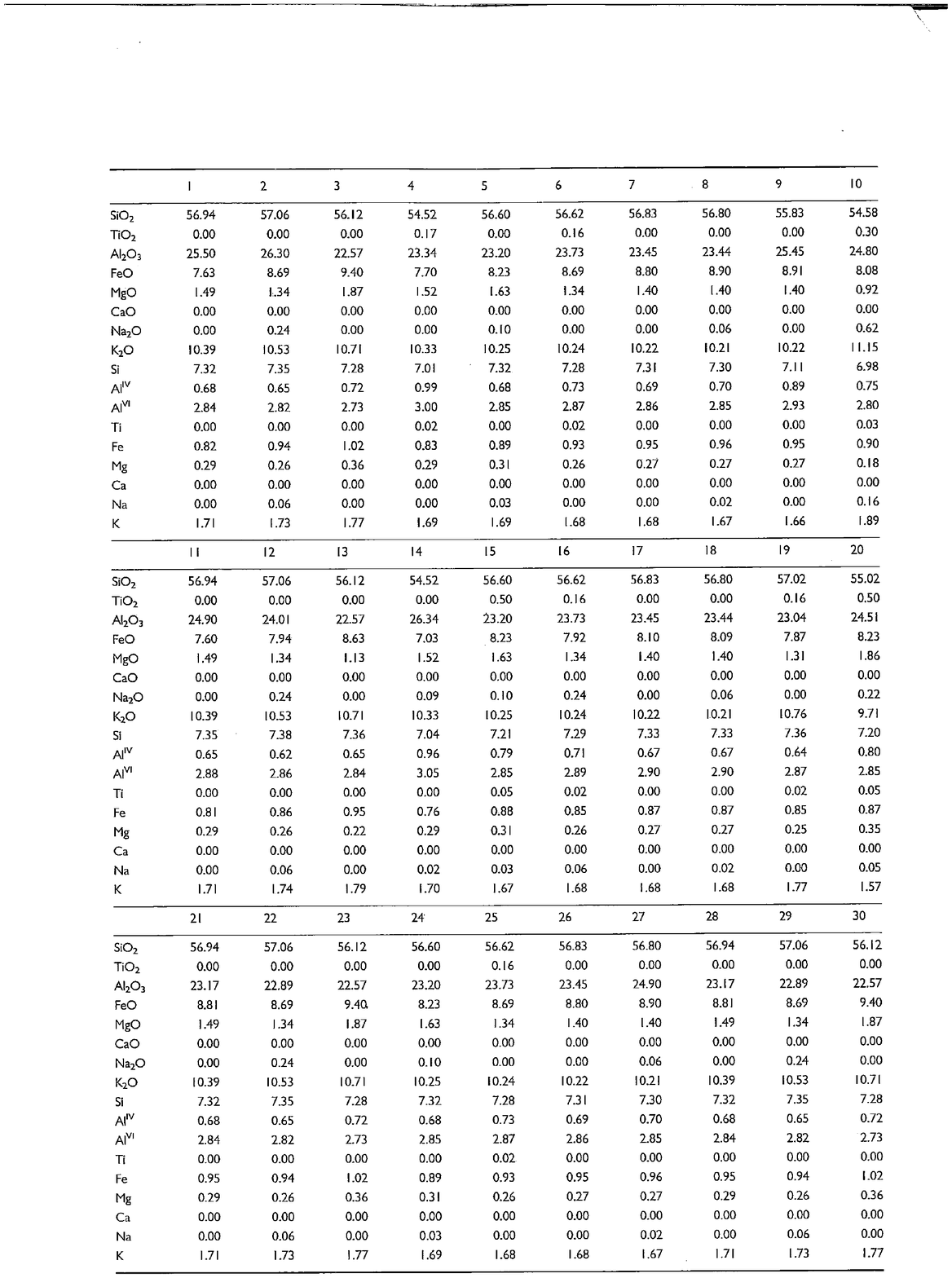}
\vspace{-2.5cm}
\caption{\label{Table I} Results of chemical analysis (wt $\%$) and atomic proportions (basis 22 oxygens) for phengite on external surface sites  \cite{Carpinteri2011}, labeled from 1 to 30.} 
\end{table}

%: Table II
\begin{table}
\vspace{-3.0cm}
\includegraphics[width=19cm]{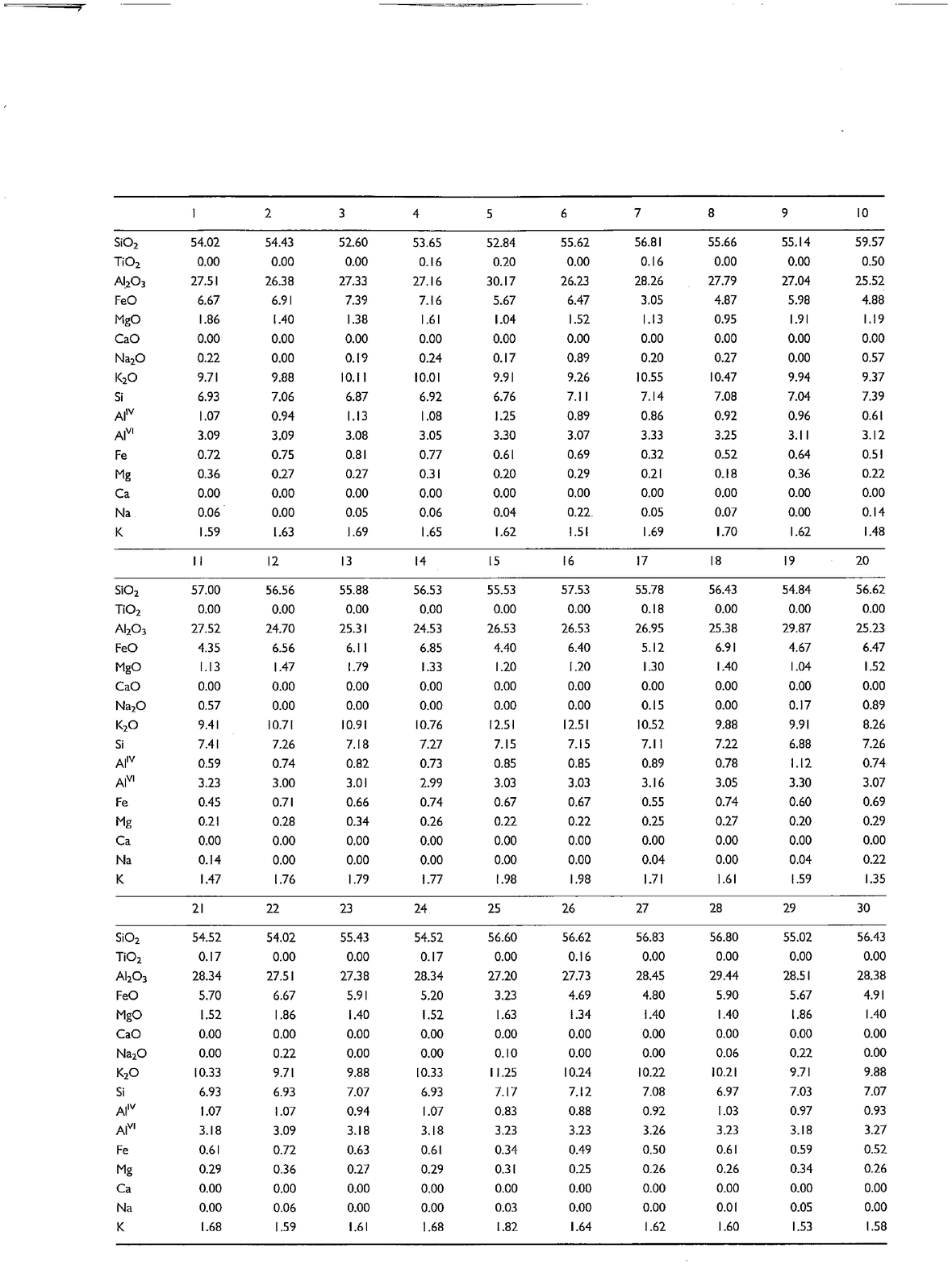}
\vspace{-2.5cm}
\caption{\label{Table II} Results of chemical analysis (wt $\%$)  and atomic proportions (basis 22 oxygens) for phengite on fracture surface sites \cite{Carpinteri2011}, labeled from 1 to 30.} 
\end{table}

%: Table III
\begin{table}
\includegraphics[width=19cm]{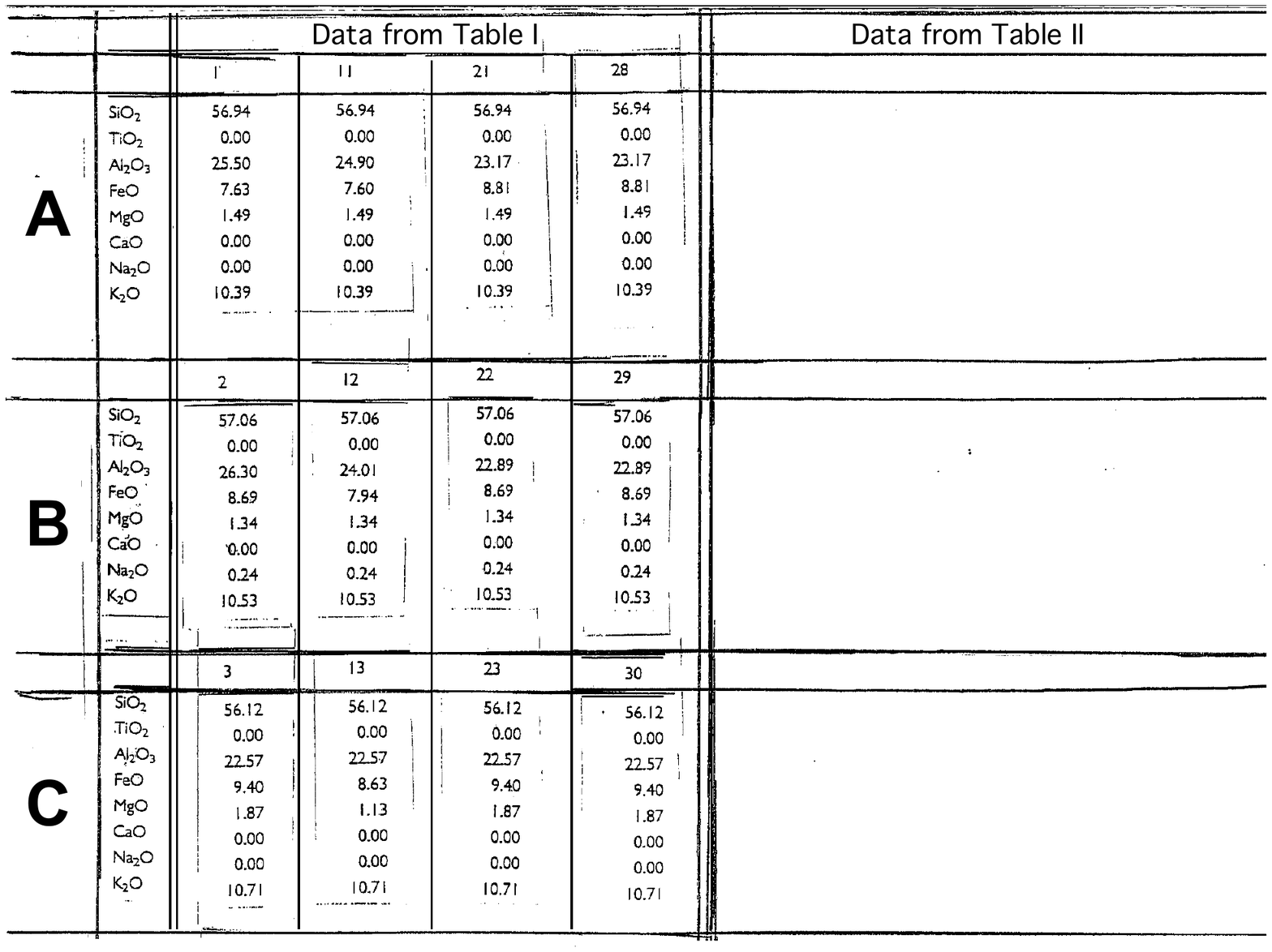}
\vspace{-13.0cm}
\caption{\label{Table III} Result of reordering of data sets contained in Tables I and II. Correlated data appear on a single table row. Left side: external surface data sets. Right side: fracture surface data sets (no data of this type appear in this table).} 
\end{table}

%: Table IV
\begin{table}
\includegraphics[width=19cm]{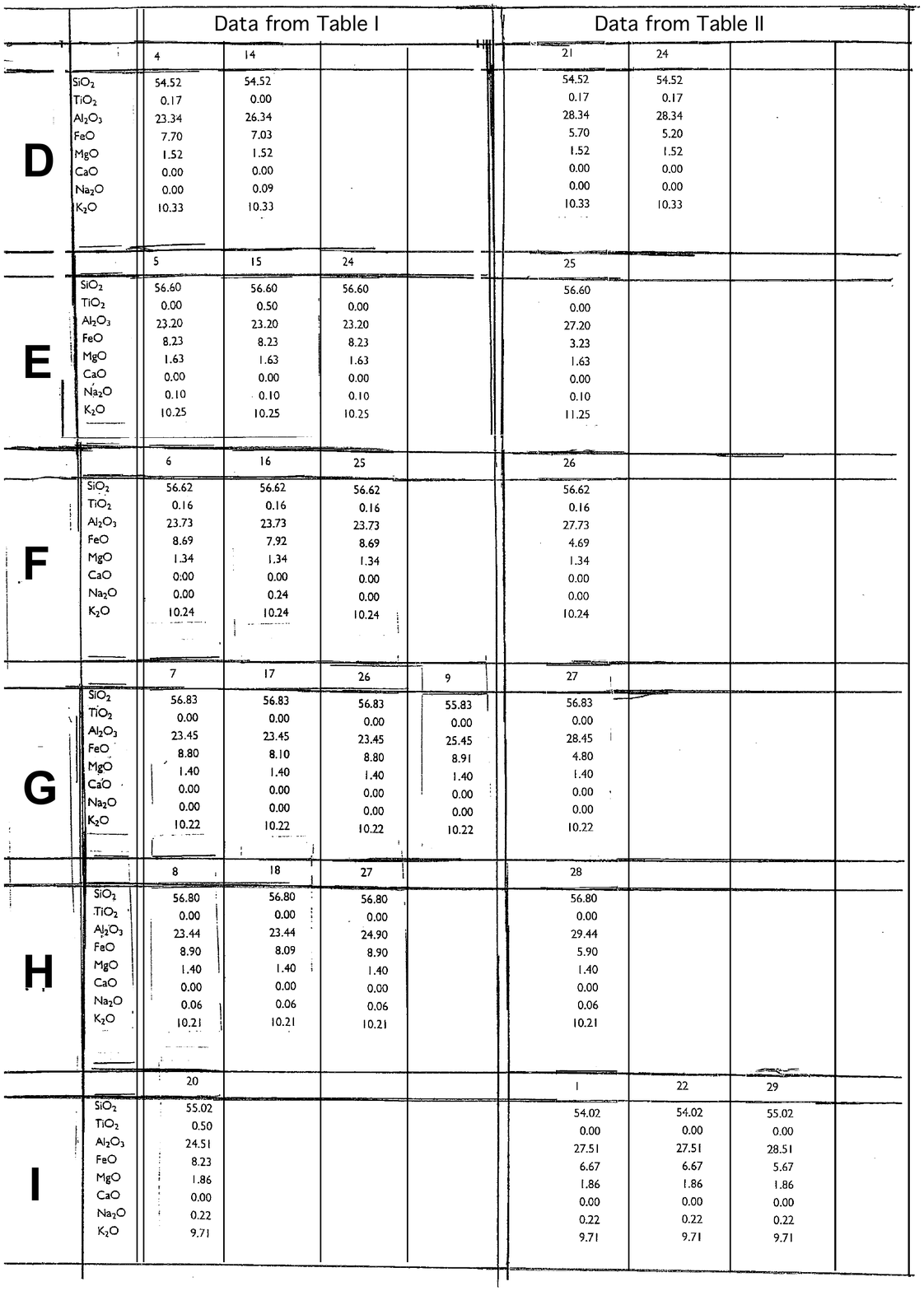}
\vspace{-2.0cm}
\caption{\label{Table IV} Result of reordering of data sets contained in Tables I and II. Correlated data appear on a single table row. Left side: external surface data sets. Right side: fracture surface data sets.} 
\end{table}

%: Table V
\begin{table}
\includegraphics[width=19cm]{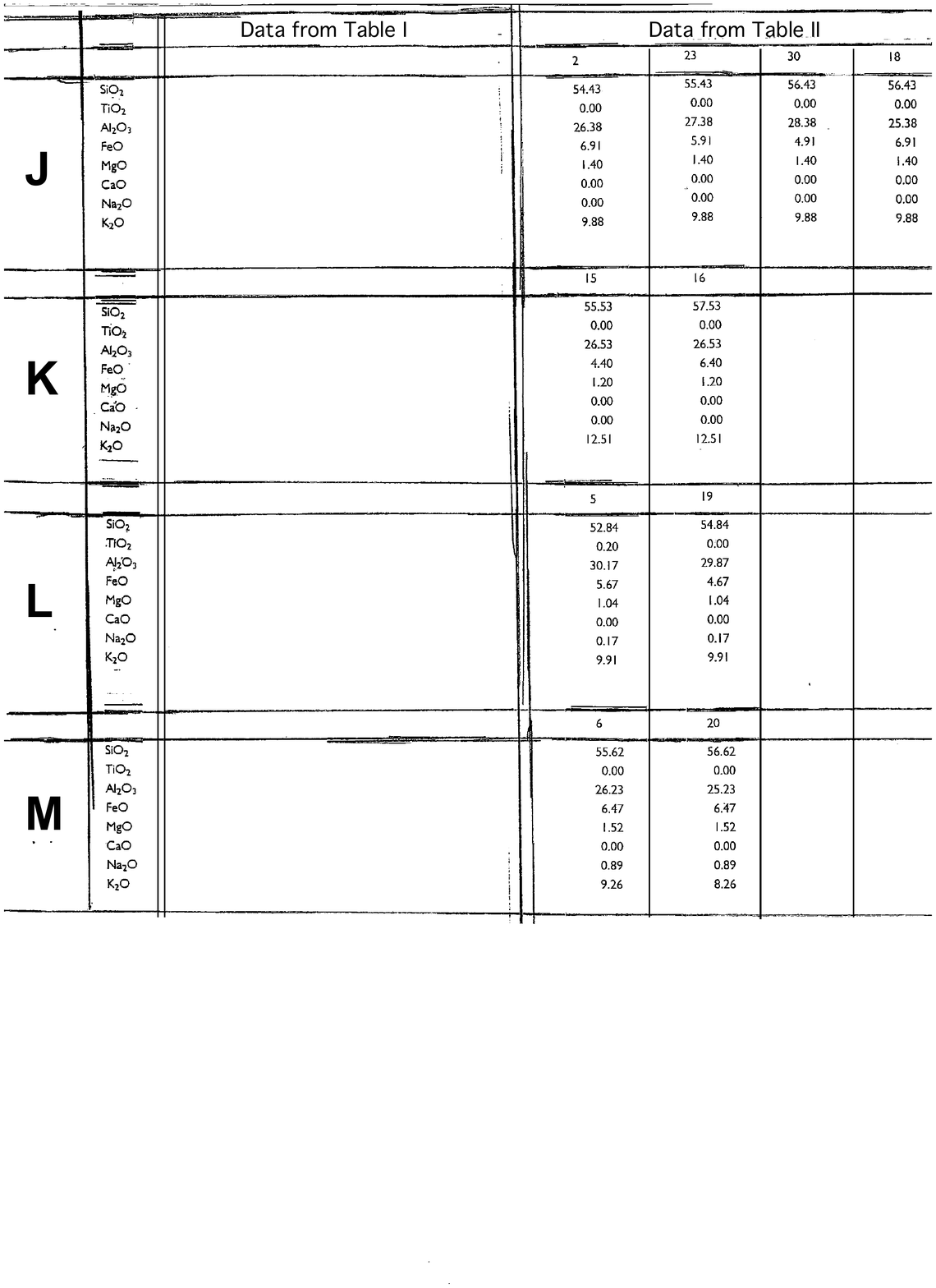}
\vspace{-9.0cm}
\caption{\label{Table V} Result of reordering of data sets contained in Tables I and II. Correlated data appear on a single table row. Left side: external surface data sets (no data of this type appear in this table). Right side: fracture surface data sets.} 
\end{table}

%: Table VI
\begin{table}
\includegraphics[width=19cm]{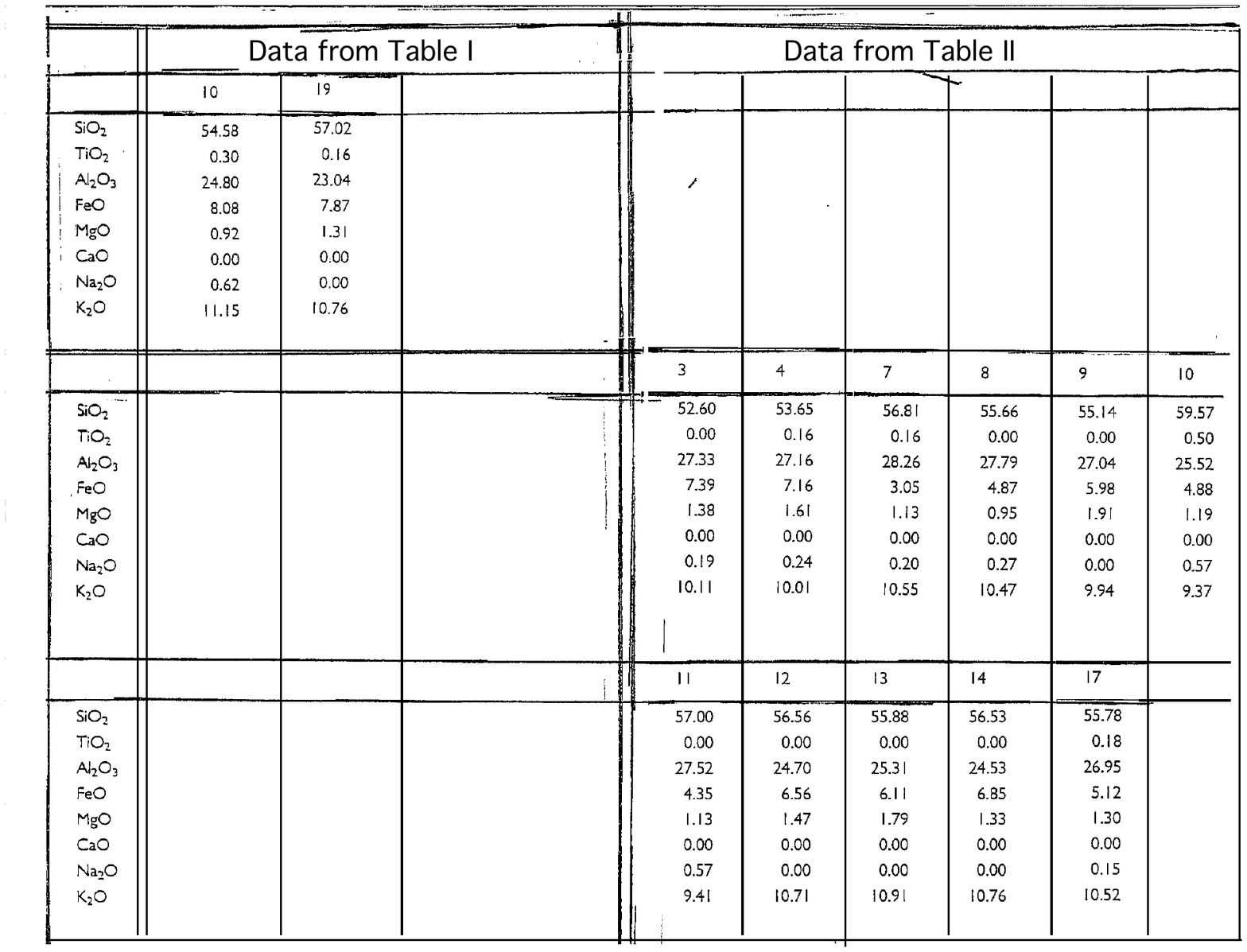}
\vspace{-13.0cm}
\caption{\label{Table VI} Data sets from Tables I and II for which no evident correlation with other data sets could be identified.} 
\end{table}

%\end{multicols}

%\bibliography{hyst}
%\bibliographystyle{prsty}

\end{document}